\documentclass{spie}  

\usepackage{amsmath,amsfonts,amssymb}
\usepackage{graphicx}
\usepackage[colorlinks=true, allcolors=blue]{hyperref}

\usepackage{gensymb}

\title{Short life and abrupt death of PicSat, a small 3U CubeSat dreaming of exoplanet detection}

\author[a]{M. Nowak}
\author[a]{S. Lacour}
\author[a]{A. Crouzier}
\author[a]{L. David}
\author[a]{V. Lapeyr\`ere}
\author[a]{G. Schworer}
\affil[a]{LESIA, Observatoire de Paris, Universit\'e PSL, CNRS, Sorbonne Universit\'e, Univ. Paris Diderot, Sorbonne Paris Cit\'e, 5 place Jules Janssen, 92195 Meudon, France}

\authorinfo{E-mail: mnowak@obspm.fr}

\begin{document} 
\maketitle

\begin{abstract}
  PicSat was a three unit CubeSat (measuring $30\textrm{ cm}\times{}10\textrm{ cm}\times{}10\textrm{ cm}$) which was developed to monitor the $\beta$ Pictoris system. The main science objective was the detection of a possible transit of the giant planet $\beta$ Pictoris b's Hill sphere. Secondary objectives included studying the circumstellar disk, and detecting exocomets in the visible band. The mission also had a technical objective: demonstrate our ability to inject starlight in a single mode fiber, on a small satellite platform.
  To answer all those objectives, a dedicated opto-mechanical payload was built, and integrated in a commercial 3U platform, along with a commercial ADCS (Attitude Determination and Control System). The satellite successfully reached Low Earth Orbit on the PSLV-C40 rocket, on January, 12, 2018. Unfortunately, on March, 20, 2018, after 10 weeks of operations, the satellite fell silent, and the mission came to an early end. Furthermore, due to a failure of the ADCS, the satellite never actually pointed toward its target star during the 10 weeks of operations.
  In this paper, we report on the PicSat mission development process, and on the reasons why it did not deliver any useful astronomical data.
\end{abstract}


\section{Introduction}
Beta Pictoris is a young ($< 20\textrm{ Myr}$), bright ($M_V = 3.86$) star system, which has been extensively studied over the past decade. We know today that this stellar system hosts a giant planet, $\beta$ Pictoris b \cite{Lagrange2009}, which is thought to be on a close to transiting orbit \cite{Lecavelier2016, Wang2016}. Although it seems unlikely that the planet itself could transit in front of its star, a transit of its Hill sphere is possible, sometime between summer 2017 and summer 2018. This event is an excellent opportunity to study the close environment of this young giant planet, but a constant monitoring of the system for such an extended period of time is difficult to set up.

A few projects have been designed to monitor $\beta$ Pictoris, including bRING \cite{Stuik2017}, ASTEP \cite{Mekarnia2017}, and observations with the BRITE constellation \cite{Weiss2014}. Three years ago, we also proposed to develop and launch PicSat, a three unit CubeSat (3U CubeSat) with the aim of detecting the Hill sphere transit, as well as to study the disk and observe exocomets in the visible band.

PicSat was successfully launched on the PSLV-C40 on January, 12, 2018, but due to a platform malfunction, the mission came to an early end on March, 20, 2018. In this paper, we report on the mission design process, and on the short 10 weeks of mission operation.

\section{Mission design overview}

Early-on in the development process of PicSat, it was decided that the entire satellite bus (structure, power system, communication system, solar panels), as well as the ADCS (Attitude Determination and Control System), would be made from existing COTS components (Commercial Off-The-Shelf components). Only the development of the science payload, the flight software, the ground segment, and the AIT activities (Assembly, Integration and Tests) were to be performed in-house at the Paris Observatory.

\subsection{Science payload}

\begin{figure}
  \begin{center}
    \includegraphics[width=0.8\linewidth]{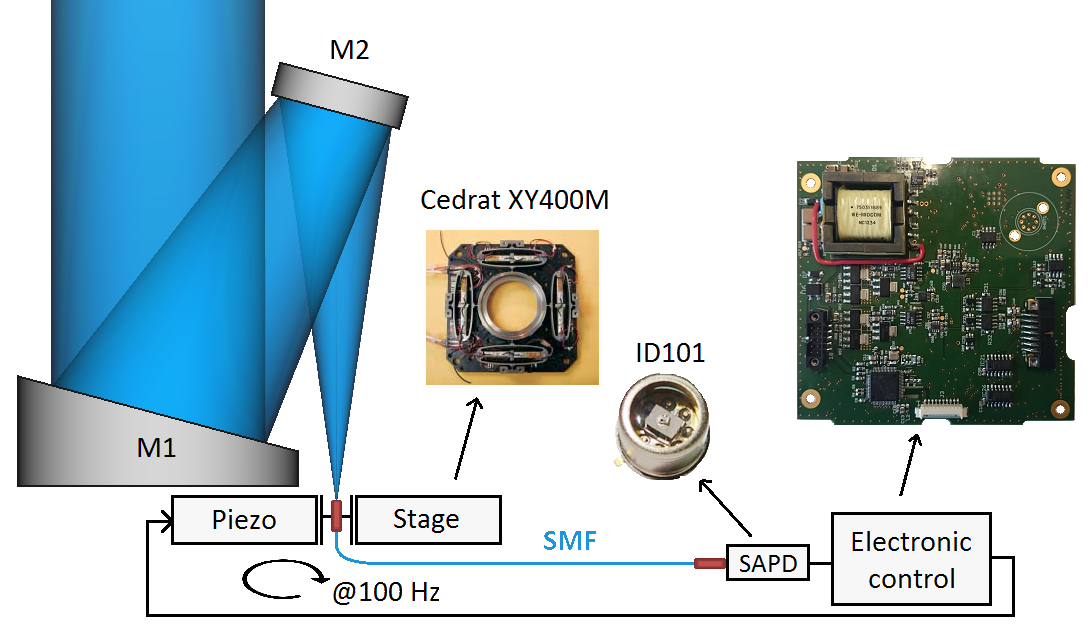}
    \caption{The general concept of the Fibered Photometer on-board PicSat.}
    \label{fig:payload_concept}
  \end{center}
\end{figure}

One of the secondary objective of the PicSat mission was to demonstrate our ability to inject starlight in a single mode fiber on a CubeSat platform, as a first step towards future CubeSat interferometers\cite{Lacour2014}. To combine this technical objective with our $\beta$ Pictoris primary science objective, we developed a small fibered photometer. The general concept is illustrated in Figure~\ref{fig:payload_concept}. In this concept, a small off-axis optical telescope (35 mm effective diameter, F/D = 4) injects the starlight into a single mode optical fiber, which brings the photons to a SPAD (Single Photon Avalanche Photodiode). To compensate for possible pointing errors of the satellite platform, the head of the optical fiber is mounted on a two-axis piezo stage, and a dedicated algorithm controls its XY position in real time (at 1 kHz). The detector is controlled in temperature by a thermo-electric module.

The entire payload (detector acquisition and temperature regulation, fiber positioning, house-keeping, data collection, etc.) is controlled by a dedicated electronic board, based on a small STM32F3 microcontroller.

This concept has four major advantages for doing photometry on a CubeSat platform:
\begin{itemize}
\item[\textbullet]{using an off-axis parabola makes for a compact optical setup, well-suited for a small CubeSat platform}
\item[\textbullet]{a single mode fiber provides extreme spatial filtering, and is largely insensitive to scattered light (from the Earth, the Moon, or the Sun), thus alleviating the need for a bulky baffle}
\item[\textbullet]{the SPAD has no readout noise, which makes possible to use very short integration times (1 ms), and opens up the possibility to use the science detector as a sensor for a fine-pointing control loop (CubeSat ADCS providers are making progress, but for the time being, pointing errors of at least 30 arcseconds are to be expected on this type of platforms)}
\item[\textbullet]{the overall mass and power budgets (1.3 kg, 1.5 W) is perfectly suited for a CubeSat mission}
\end{itemize}  

\begin{figure}
  \begin{center}
    \includegraphics[width=0.4\linewidth]{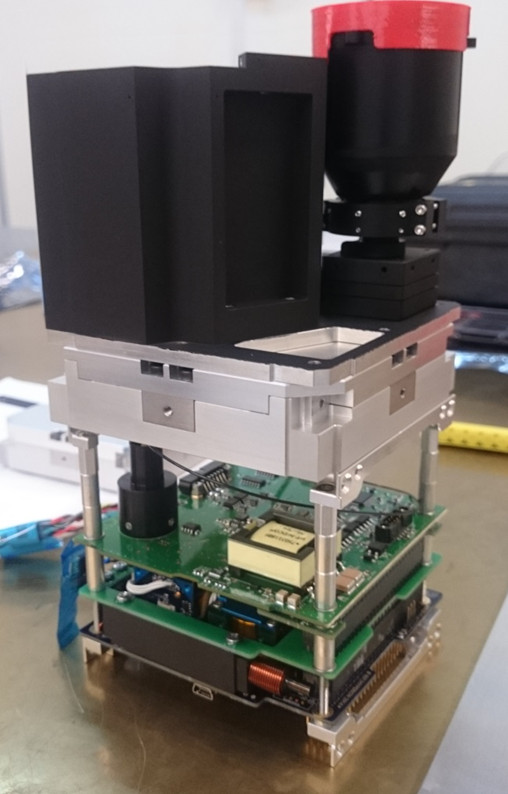}
    \caption{A view of the flight model of the PicSat payload, fully integrated. From bottom to top: ADCS electronic board (not part of the payload), payload electronic board (with the SPAD in black), two-axis piezo stage (in aluminum gray), and the optical telescope (top left), next to the star tracker (top right, with a red cap, not part of the payload).}
    \label{fig:payload_fm}
  \end{center}
\end{figure}

The payload (engineering model) was tested in thermal vacuum, vibrations, and shocks. The flight model was integrated and aligned in August 2017 at the Paris Observatory (see Fig. \ref{fig:payload_fm}).

\subsection{Satellite platform}

The PicSat platform is a 3U CubeSat platform, made by the Dutch company ISIS (Innovative Solutions In Space\footnote{www.isispace.nl}). They provided the satellite structure, the On-Board Computer, the Electrical and Power System (subcontracted to GOMSpace), and the solar panels.

For the payload fine pointing system to be able to lock the fiber onto the star, the main ADCS needs to provide at least a 1 arcmin pointing precision (at 1 sigma). The ADCS selected for the PicSat Mission is the iADCS 100, made by Hyperion Technologies\footnote{www.hyperiontechnologies.nl}, also a company based in the Netherlands. This system features three reaction wheels, three magnetorquers, and it includes their ST200 star tracker as the main sensor for precision target pointing.

\begin{figure}
  \begin{center}
    \includegraphics[width=0.8\linewidth]{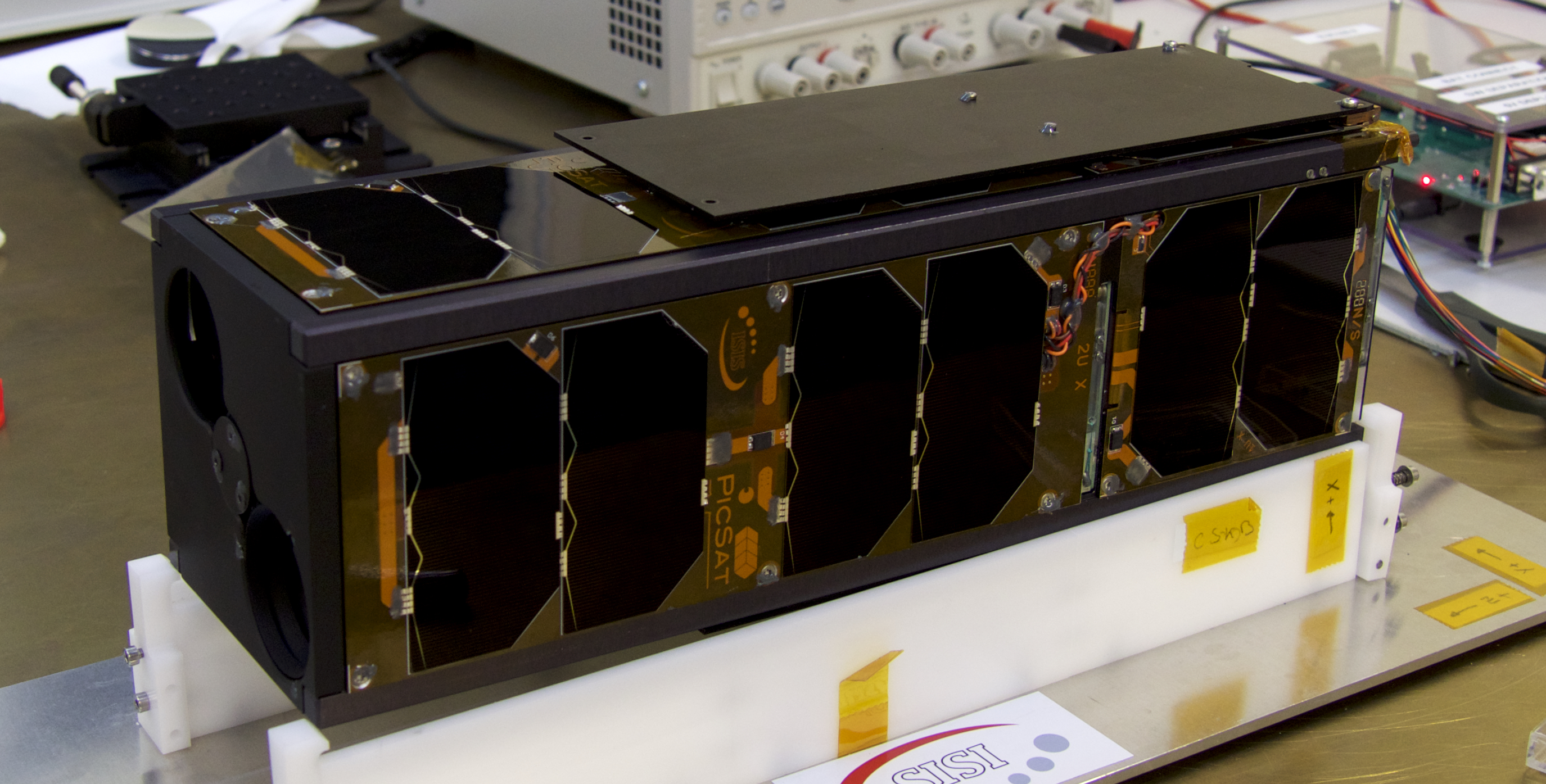}
    \caption{A view of PicSat (flight model) fully integrated and ready to be delivered to the launch provider.}
    \label{fig:picsat_fm}
  \end{center}
\end{figure}

A view of the fully integrated flight model is given in Fig~\ref{fig:picsat_fm}.

\subsection{Ground segment}

\begin{figure}
  \begin{center}
    \includegraphics[width=0.8\linewidth]{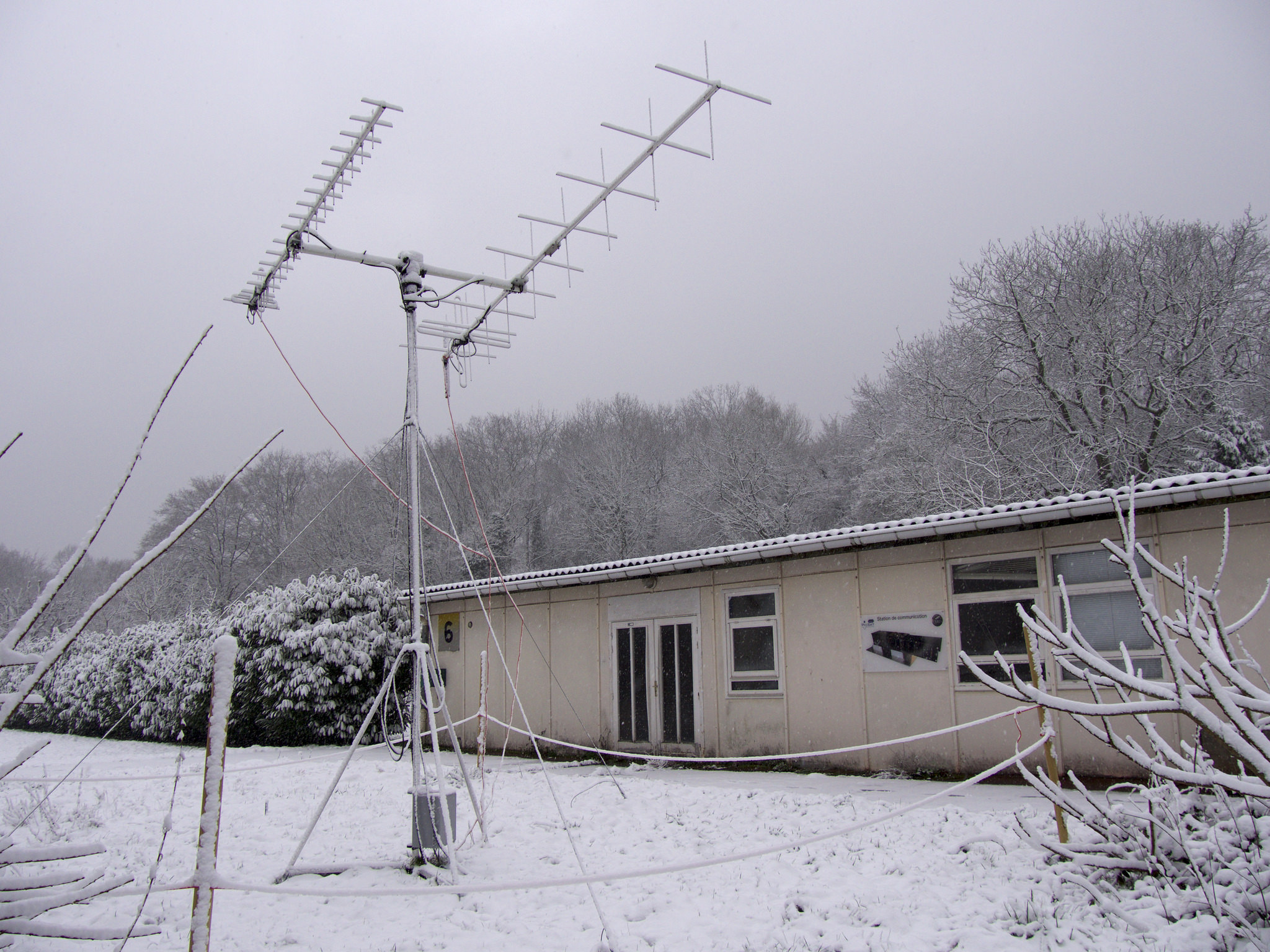}
    \caption{A view of the UHF/VHF antenna of the PicSat Operation Center, in Meudon, France.}
    \label{fig:antenna}
  \end{center}
\end{figure}

The satellite communicates with the ground using its two UHF/VHF antennas (UHF downlink, VHF uplink). All mission operations are performed from our operation center, in Meudon, France, using a dedicated UHF/VHF Yagi antenna (Fig.~\ref{fig:antenna}). In downlink, the antenna is connected to a SDR (Software Defined Radio) dongle. In uplink, data packets are sent to the antenna through a Kenwood TS-2000E TNC. Control of the antenna (to track the satellite when passing over Meudon) and of the TNC frequency (to account for Doppler shift) is done using the GPredict free software\footnote{http://gpredict.oz9aec.net}.

All the software required to handle data packets and store in/retrieve from the mission database has been developed in-house, in Python. The software used to demodulate the incoming data (downlink) is PicTalk\footnote{https://github.com/f4gkr/PicTalk}, which was developed for this mission by F4GKR, but which can potentially be used for other missions.


\subsection{Development timeline and project cost}

The idea of building a small CubeSat to monitor $\beta$ Pictoris and study the transit of the giant planet $\beta$ Pictoris b emerged in 2014, and activity on the project started in early 2015, less than 3 years before the expected time of transit. The overall mission development schedule was constrained by the need to have the satellite in orbit as soon as possible (mid 2017 best case, early 2018 worst case).

\begin{figure}
  \begin{center}
    \includegraphics[width=0.8\linewidth]{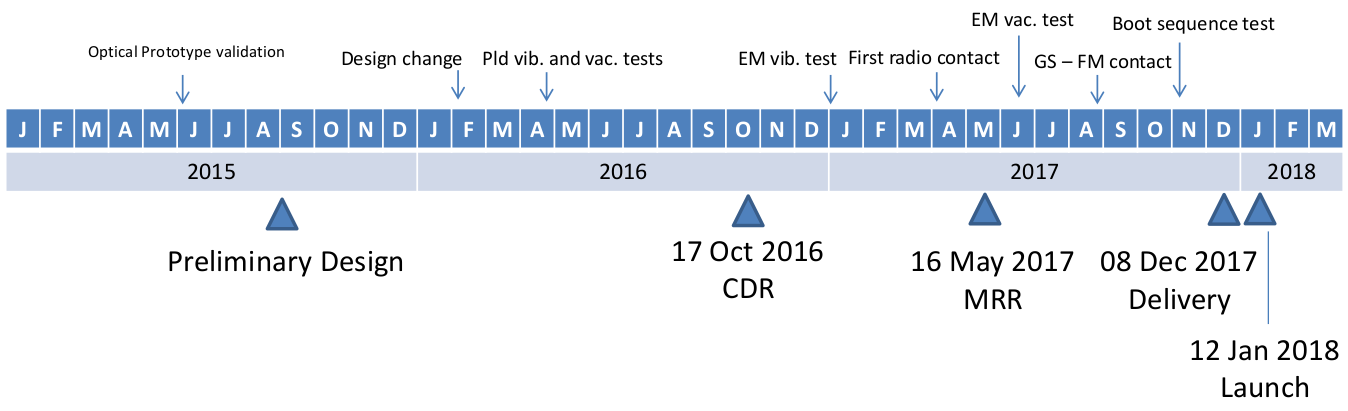}\\[0.5cm]
    \includegraphics[width=0.7\linewidth, trim=0cm 0cm 5cm 0cm, clip = True]{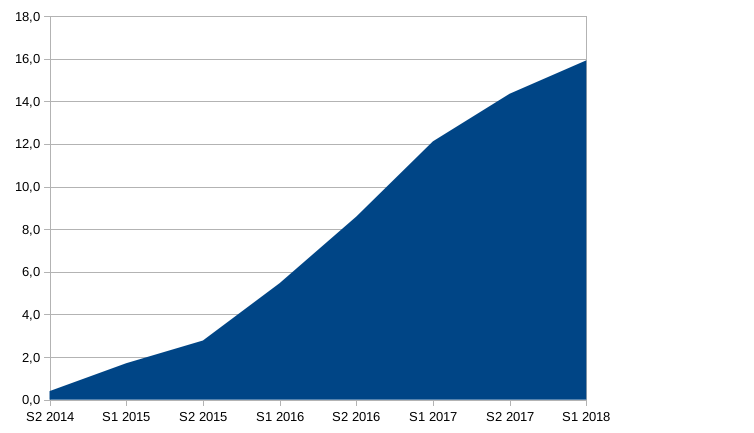}\\[0.5cm]
    \includegraphics[width=0.7\linewidth, trim=0cm 0cm 5cm 0cm, clip = True]{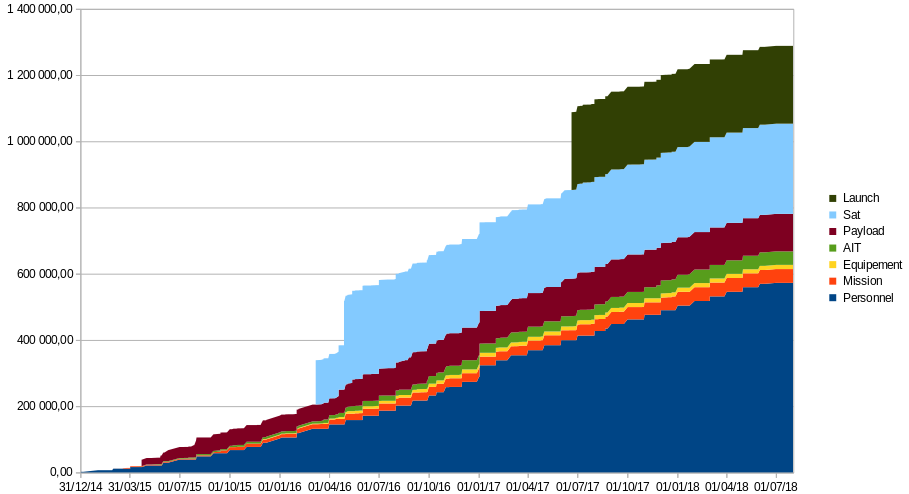}\\
    \caption{A summary of the development timeline of the project (top panel), along with the cumulative number of FTEs (middle panel), and the budget (bottom panel). The cumulative budget is given in euros, and is separated in salaries (blue), travels (orange), lab equipment (yellow), AIT activities (green), payload hardware (red), satellite hardware (light blue), launch (dark green).}
    \label{fig:project}
  \end{center}
\end{figure}

The general timeline is given in Figure~\ref{fig:project}, along with the cumulative number of FTEs (Full-Time Equivalents), and the cumulative budget.

In summer 2015, a first prototype of the payload was built and tested, to validate the concept of doing photometry with a single mode fiber. Early environmental tests on the payload were performed in December 2015, and revealed a problem with the piezo actuator, leading to a new and slightly bigger design of the payload. This new design was quickly assembled to be tested again in vibration and thermal vacuum, in order to validate it, and to move forward with the assembly of the complete engineering model. In October 2016, about a year after the project had started, we passed a Critical Design Review, and two months later performed a full engineering model vibration test. This validated the final design of PicSat, and we moved on to the assembly of the flight model. We also started to work on the ground segment (hardware and software) in autumn 2017. A Mission Readiness Review was passed in mid 2017. At the time, the mission was clearly not ready to fly, but the review helped to clarify where to put the most effort in order to have a satellite ready to be delivered in December 2017. The satellite was successfully delivered on December, 8, 2017.

\section{Launch, early operations, and mission failure}

\subsection{First contact}

After its delivery to ISL (Innovative Space Logistics\footnote{https://www.isispace.nl/launch-services}), in the Netherlands, the satellite was sent to India, integrated on the PSLV-C40, and launched from  Sriharikota on January, 12, 2018, at 09:28 am local time (03:28 UTC). The launch was successful, and the satellite was deployed on a 505 km Sun-Synchronous orbit.

On its 505 km orbit, the satellite was passing over our ground station in Meudon, France, about 4 to 6 times per day (2 to 3 passes in the morning, and 2 to 3 in the evening). Each pass lasted between 5 to 15 minutes (from horizon to horizon), with around 5 to 10 minutes of useful communication time. We received the satellite beacon (a data packet emitted automatically every 10 s) on the first pass of the satellite over our station, and we decoded our first packets the same day, during the evening pass. The second day, we sent the first telecommand, and received the first answer from the satellite, confirming the proper functioning of our communication systems. This was the beginning of about 10 weeks of PicSat operations.

\begin{figure}
  \begin{center}
    \includegraphics[width=0.9\linewidth]{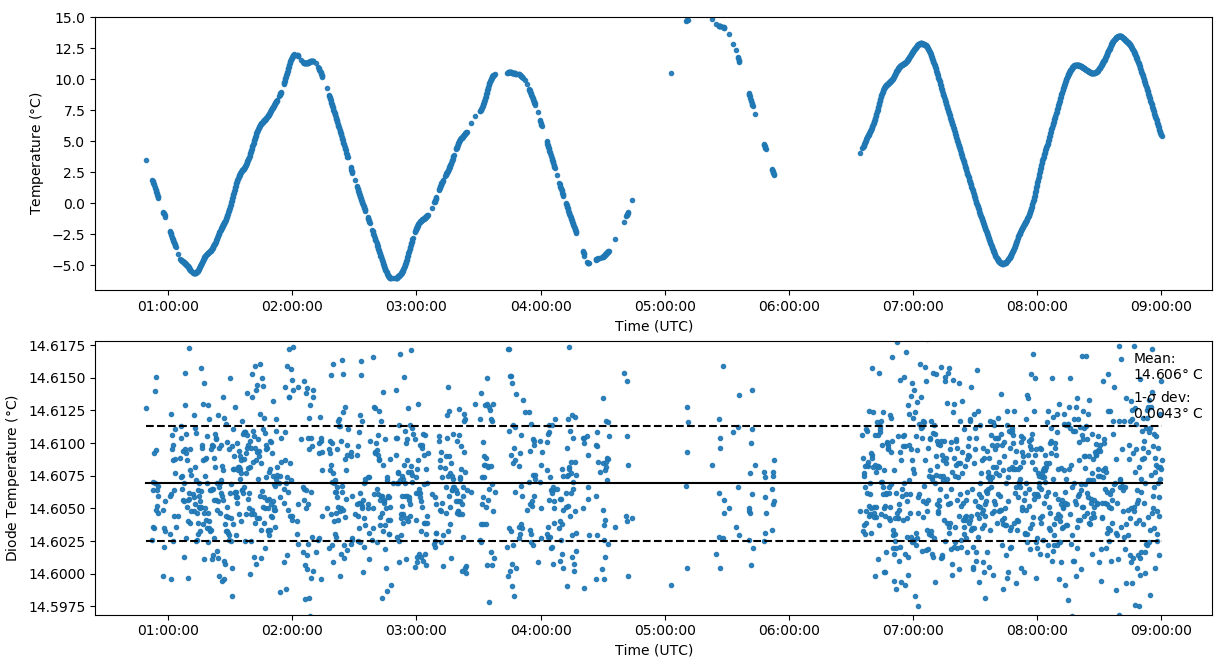}
    \caption{Evolution of the temperature of the payload electronic board (top panel), and of the temperature of the detector (bottom panel), on five consecutive orbital cycles.}
    \label{fig:payload_temperatures}
  \end{center}
\end{figure}

About a week after the launch, we successfully deployed the two deployable solar panels, and two weeks after launch, we activated the science payload, and confirmed that the electronic board was working properly. We also confirmed that the temperature regulation of the detector was working, and that the temperature of the detector was stable at $14.606\pm{}0.004\degree\mathrm{C}$ (see Fig.~\ref{fig:payload_temperatures})

\subsection{ADCS failure}

Two critical tasks were devoted to the iADCS 100 on-board PicSat. The first task was to de-tumble the satellite after its initial release from the launcher, to allow the star tracker to get a fix (maximum slew rate tolerated by the ST200 is $0.3\textrm{ deg/s}$ in tip/tilt, $0.6\textrm{ deg/s}$ in roll, as per datasheet specification). Then, the ADCS was supposed to point the satellite toward $\beta$ Pictoris, its target star, and maintain this inertial pointing for the entire duration of the mission (1 yr, nominal).

\begin{figure}
  \begin{center}
    \includegraphics[width=0.9\linewidth]{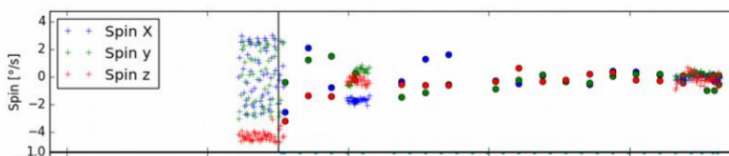}
    \caption{Evolution of the rotation rate on the 3 axes of the satellite measured by the ADCS internal sensors, before and after initiating the de-tumbling mode on the iADCS 100 (start de-tumbling is represented by the vertical black line).}
    \label{fig:detumbling}
  \end{center}
\end{figure}

Two weeks after launch, the command to activate the de-tumbling was sent to the satellite, with the aim of switching to target pointing and starting science activities soon after. In a matter of a few orbits, the rotation rate of the satellite was brought from a few degrees per second to about $0.1\textrm{ deg/s}$ (Fig.~\ref{fig:detumbling}). 

\begin{figure}
  \begin{center}
    \includegraphics[width=0.9\linewidth]{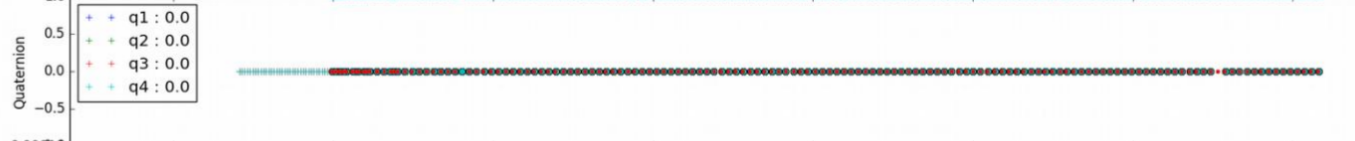}
    \caption{One hour of quaternion data returned by the iADCS100.}
    \label{fig:zeroquaternion}
  \end{center}
\end{figure}

However, despite that this rotation rate was well below the maximum tolerated slew rate of the ST200, the system did not return any quaternion when switching to the target pointing mode (Fig. \ref{fig:zeroquaternion}). For the next 8 weeks, we worked with Hyperion Technologies to try to solve the issue. We confirmed that the ST200 was alive, as were all the independent subsystems of the iADCS100. But there seemed to be some high-level software issues with the system. Despite 8 weeks of tedious in-orbit debugging, we never managed to point to the science target.

\subsection{End of mission}

\begin{figure}
  \begin{center}
    \includegraphics[width=\linewidth, trim = 2cm 0cm 2cm 0cm, clip = True]{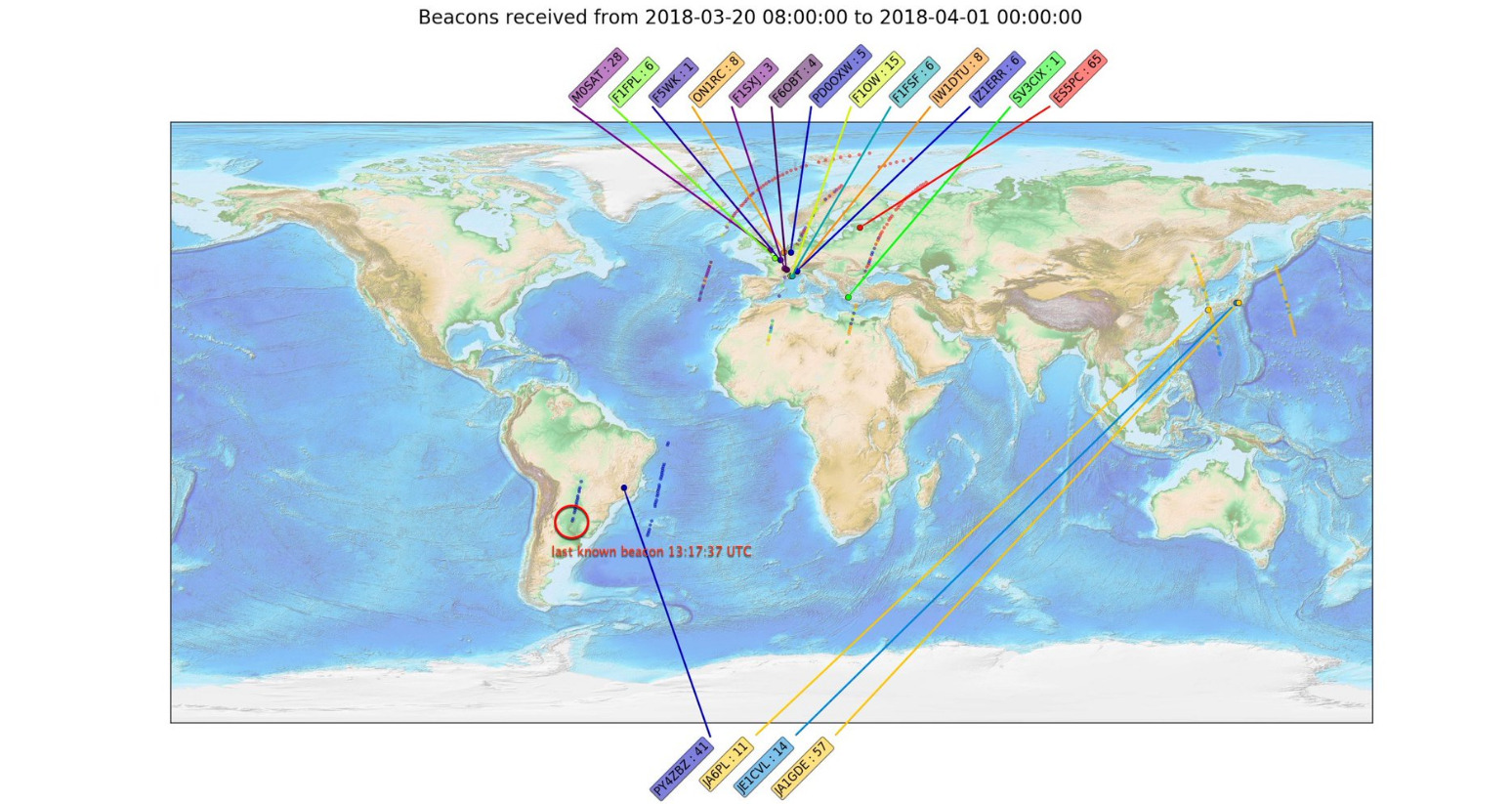}
    \caption{Map of the last beacons received by all radioamateurs listening to PicSat during the last few orbits before end-of-mission}
    \label{fig:last_beacon}
  \end{center}
\end{figure}

While we were still working on trying to get the ADCS to point the satellite towards $\beta$ Pictoris, after about 10 weeks of operations, the satellite suddenly fell silent, on March, 20, 2018. The satellite was emitting its radio beacon every 10 s on amateur frequencies, and we had a network of 80 radioamateurs following it every day. Thus, our ground coverage was extensive, and we managed to pinpoint the time and location at which PicSat stopped emitting. In Fig.~\ref{fig:last_beacon}, we show a map of the last few orbits of the satellite, with all the beacons received around the world. The last beacon was received above South America, at 13:17 UTC. After that, the satellite was supposed to fly over the South pole, and then over Indonesia, where amateurs reported not receiving it. We concluded that PicSat stopped working sometime between 13:17 UTC and 13:47 UTC on March 20, while flying over the South pole, between South America and Indonesia.

An extensive review of all the telemetries received before the end of mission revealed no particular anomaly, and no warning sign. The satellite did experience a rather large number of reboots (about 1 reboot of the On-Board Computer or of the Electrical Power System every 3 days), but the frequency did not change much throughout the mission. All temperatures, voltages, and status were nominal, even in the last beacon received.

We also want to point out that during the last few passes of the satellite over our station in Meudon, we were performing a software update. The on-board software of PicSat had a dual-layer architecture. A Level-0 software was in charge of all the critical tasks (power, communication, de-tumbling), whereas a second layer (Level-1), which was only loaded on command, was in charge of mission specific non-critical tasks (mainly managing the payload, and science data processing). The Level-0 software could not be changed on-orbit, but could be used to update the Level-1.
During the last pass of the satellite over Meudon, after having spent several days sending pieces of the new Level-1 software to the satellite, we requested the Level-0 to rewrite the Level-1. However, the new Level-1 software loaded on-board was found to be corrupted, and the command was rejected, meaning that something went wrong during the upload of the new software on the satellite.

Given all this, we had three main hypotheses to explain this early end-of-mission:
\begin{itemize}
\item[\textbullet]{A malfunction of the downlink power amplifier of the communication board (this happened on the ground)}
\item[\textbullet]{A software bug, leading to a bootloop of the system}
\item[\textbullet]{Given the location of the loss of contact, a radiation/charged particle damage on one of the numerous critical electronics (main computer, power system, communication system) is also a possible explanation.}
\end{itemize}

The first of these three hypotheses was ruled out by listening to the satellite using the 25 m dish antenna at Dwingeloo (which gives a 40 dB gain, enough to compensate for a defective power amplifier).

The last two hypotheses are still plausible. An extensive investigation of a possible software bug led us to the discovery of a vulnerability in our on-board software updater, which can indeed lead to a system bootloop (upon reboot, the Level-0 would try to execute a sequence of operations which would lead to a memory corruption, triggering a reboot, and thus the execution of the same sequence over and over again). In this case, the computer would reboot before even switching on the communication system, thus effectively turning the satellite into a piece of space junk. However, we note that despite all our greatest efforts, we never managed to trigger this bootloop on the engineering model, except when sending sequences of command specifically engineered to exploit the vulnerability. In particular, we reproduced the entire sequence of commands used for the software update (which had already been tested on the engineering model before trying to update the flight model), and did not trigger a bootloop.

\begin{figure}
  \begin{center}
    \includegraphics[width=0.7\linewidth]{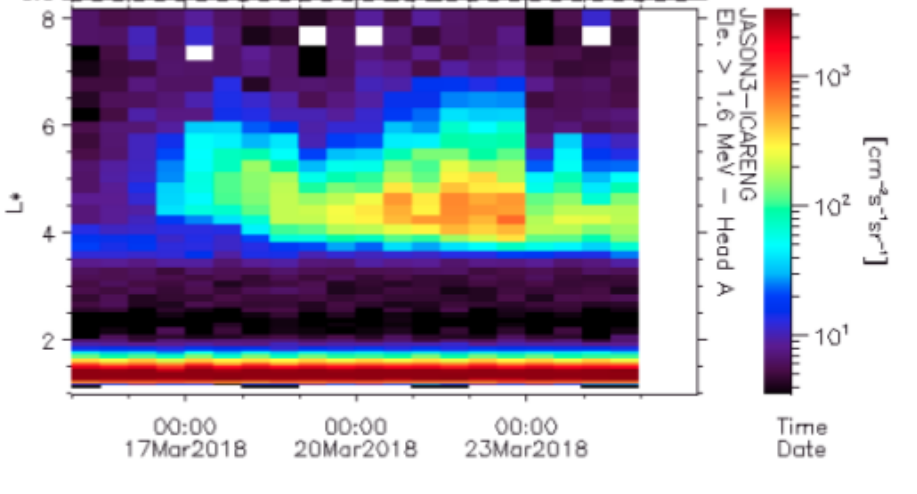}
    \caption{Evolution of the density of high-energy electrons in the outer Van Allen belt at the time of PicSat end-of-mission (data from JASON3 ICARE-NG)}
    \label{fig:density}
  \end{center}
\end{figure}

\begin{figure}
  \begin{center}
    \includegraphics[width=0.45\linewidth]{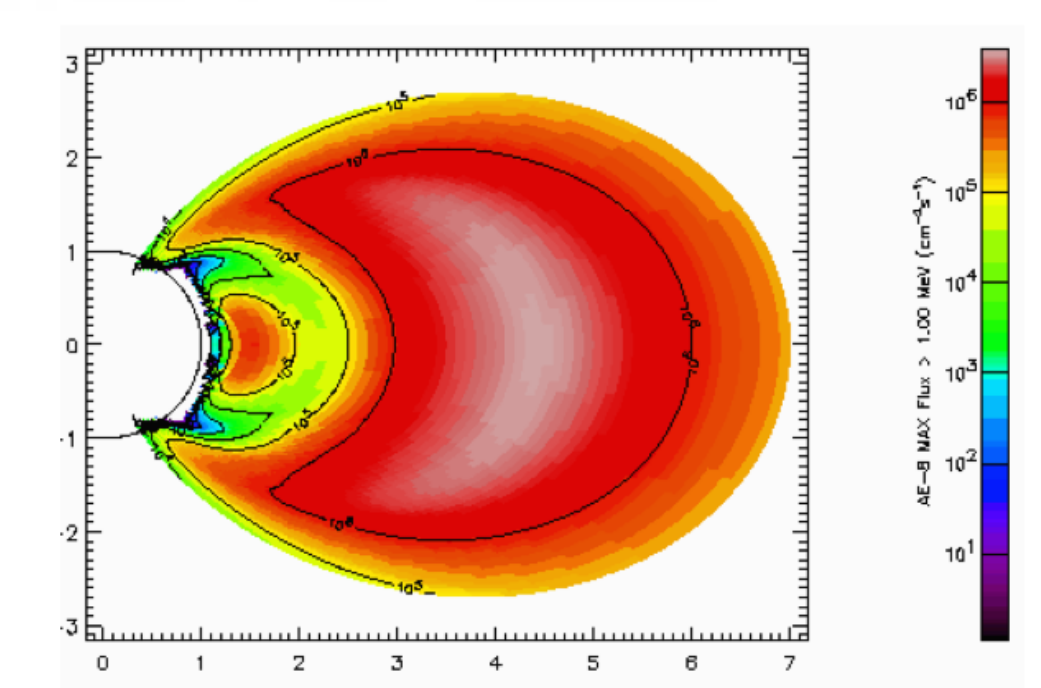}
    \includegraphics[width=0.45\linewidth]{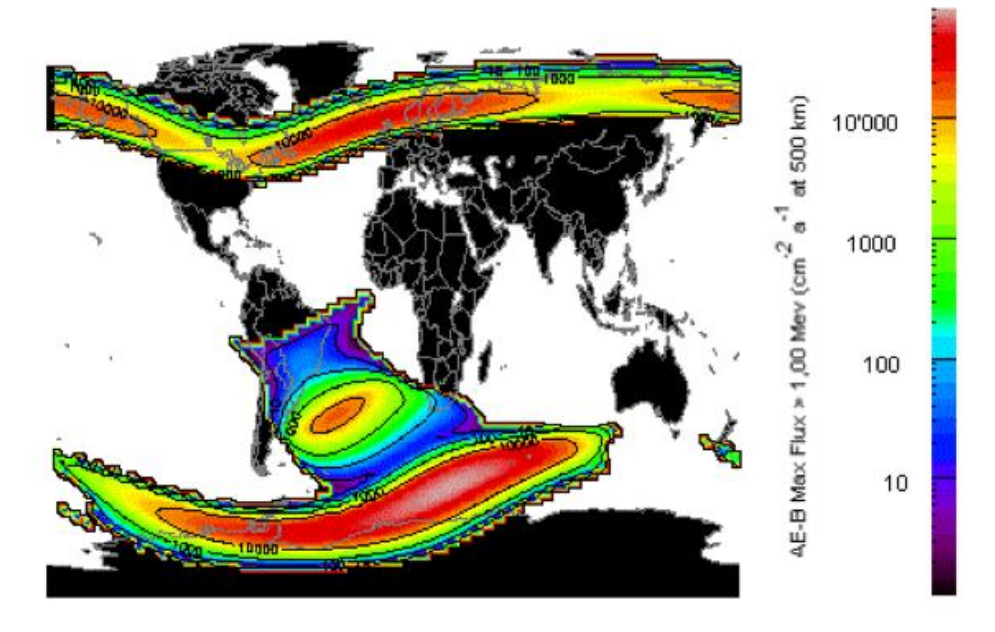}
    \caption{Map of the integral electron flux $> 1\textrm{ MeV}$  in the outer belt (left panel), and at 500 km altitude above the Earth (right panel). This does not correspond to the situation at the time of end-of-mission, but only  illustrates how the outer Van Allen belt connects to the poles. Figures from https://www.spenvis.oma.be/help/background/traprad/traprad.html, data from \cite{Vette1991}}
    \label{fig:radiations}
  \end{center}
\end{figure}

We also investigated the last hypothesis, and found that a few day before the loss of contact, there had been a magnetic event in the Earth magnetosphere, which raised the density of high-energy electrons in the outer Van Allen radiation belt. As can be seen in Fig.~\ref{fig:density}, this particle density increased from March 17 to March, 22, when it peaked, before going down to its normal level. The outer Van Allen belt is usually located at much higher altitudes (between 4 to 6 Earth radii), than Low-Earth Orbit ($< 1000\textrm{ km}$), but not at the poles, where it goes to much lower altitudes (see Fig.~\ref{fig:radiations}).

\section{Conclusion: are CubeSats and off-the-shelf components viable for astronomy?}

Scientifically speaking, the PicSat mission was a failure. It did not deliver the data it was supposed to deliver on $\beta$ Pictoris. A major malfunction of the ADCS (a commercial system) made impossible to point the target star during the 10 weeks of operations. The cause of the early end-of-life of the satellite remains unclear.

At a time when more and more CubeSat-based astronomy missions are being suggested, this raises the question of the viability of this type of platforms, and more importantly, of the COTS components approach. Other projects have demonstrated that astronomical CubeSats are possible. For example, the BRITE constellation\cite{Weiss2014}, or the ASTERIA mission\footnote{https://www.jpl.nasa.gov/cubesat/missions/asteria.php} were successful. But both of them either developed a large part of the hardware themselves, or used flight-proven components (the ASTERIA mission used the ADCS from Blue Canyon Technologies, already flight-proven). In the case of PicSat, we deliberately decided to turn to commercial hardware for most of the satellite subsystems, some of which (i.e. the ADCS) was not flight-proven.

Whether or not this was a good decision is hard to tell. This made possible to design, build, launch, and operate a mission in an extremely short time frame (3 yr), and with limited resources (16 FTEs, \$1.4 M total). But the COTS components available today for CubeSats may not have quite the level of reliability suited for a space mission. And due to the fact that most of these systems are mission critical (power, communication, main computer), it is almost impossible to know for sure which one was faulty (if any; again, this could be a software failure). This makes difficult to decide how to improve the next mission based on this return of experience. The ADCS was clearly defective. But for the next mission, should we also change the On-Board Computer? The communication board? What about adding redundancies? To which sub-systems?

Looking back at this project, we believe that more testing may be the key to mission success. As long as the reliability of commercial systems remains unclear, they should probably be tested very thoroughly. Along all the development stages of the project, we did perform several environmental tests, but only in thermal vacuum, vibrations, and shocks. We did not perform any radiation test on the electronics, mainly because these are more costly, and much more difficult to set up. We also spent a large amount of time trying to develop a dedicated test bench for the ADCS, but this turned out to be too time-consuming, and we had to give up.

As astronomers and/or instrumentalists, we wish to focus only on the development of new and ambitious payloads, able to deliver high-quality science data. But to be able do so, we need to have reliable COTS hardware for the platform, that we can safely rely on. This may not be the case, yet. Then, early testing, even when the manufacturers claim certain levels of performance/reliability for their products, is probably the only way to know for sure in which systems we can be confident, and which systems should be avoided. This may also be the only way to push manufacturers to increase the reliability of their systems.

\bibliography{picsat}
\bibliographystyle{spiebib} 

\end{document}